\documentclass [12pt,a4paper]{article}

\usepackage[english]{babel}

\usepackage[a4paper,top=2cm,bottom=2cm,left=3cm,right=3cm,marginparwidth=1.75cm]{geometry}

\usepackage{amsmath}
\usepackage{graphicx}
\usepackage[colorlinks=true, allcolors=blue]{hyperref}

\usepackage[dvipsnames]{xcolor}
\usepackage{tikz}
\usetikzlibrary{math}
\usepackage{xstring}
\usepackage{ifthen}

\usepackage{soul}
\providecommand{\keywords}[1]
{
  \small	
  \textbf{\textit{Keywords---}} #1
}

\newcommand\setAllDots[4][]{
  \foreach \a in {1,...,#2} {
    \pgfmathtruncatemacro{\jmt}{mod(\a-1,4)}
    \pgfmathtruncatemacro{\jdt}{(\a-1)/4}
    \pgfmathsetmacro\jd{-\jdt*6.-2.2}
    \pgfmathsetmacro{\jm}{\jmt*6.+1.8}
    \node[draw] at (\jm+0.5,\jd) {\a};
    \foreach \phit in {1,...,#3} {
        \pgfmathsetmacro{\phi}{(\phit-1)*360/#3 + #4}
        \filldraw[black] ({\jmt*6 + 2*cos(\phi)}, {-\jdt*6 + 2*sin(\phi)}) circle (5pt);
    }
  }
  \pgfmathtruncatemacro{\jmt}{(#2-1)/4}
  \ifthenelse{\jmt > 0}{
    \foreach \a in {1,...,\jmt} {
        \pgfmathsetmacro{\jm}{-\a*6+3}
        \draw[gray,very thin] (-2,\jm) -- (20,\jm);
    }}{}
  \foreach \a in {0,1,2} {
    \draw[gray,very thin] (\a*6+3,2.5) -- (\a*6+3,-\jmt*6-2.5);
  }
}
\newcommand\ConnN[6][]{
    \pgfmathtruncatemacro{\jmt}{6*mod(#2-1,4)}
    \pgfmathtruncatemacro{\jdt}{-6*int((#2-1)/4)}
    \pgfmathsetmacro\phiF{(#5-1)*360/#3 + #4}
    \pgfmathsetmacro\phiL{(#6-1)*360/#3 + #4}
    \draw[black,line width=1mm] ({\jmt+2*cos(\phiF)}, {\jdt+2*sin(\phiF)}) -- ({\jmt+2*cos(\phiL)}, {\jdt+2*sin(\phiL)});
}

\newcommand\colorFillDraw[3]{
    \ifthenelse{#1 = 0}{
        \filldraw[black,line width=0.3mm] ({#3*cos(#2)},{#3*sin(#2)}) circle (6pt);}{}
    \ifthenelse{#1 = 1}{
        \filldraw[Cyan,line width=0.3mm] ({#3*cos(#2)},{#3*sin(#2)}) circle (6pt);}{}
    \ifthenelse{#1 = 2}{
        \filldraw[green,line width=0.3mm] ({#3*cos(#2)},{#3*sin(#2)}) circle (6pt);}{}
    \ifthenelse{#1 = 3}{
        \filldraw[Goldenrod,line width=0.3mm] ({#3*cos(#2)},{#3*sin(#2)}) circle (6pt);}{}
    \ifthenelse{#1 = 4}{
        \filldraw[orange,line width=0.3mm] ({#3*cos(#2)},{#3*sin(#2)}) circle (6pt);}{}
    \ifthenelse{#1 = 5}{
        \filldraw[red,line width=0.3mm] ({#3*cos(#2)},{#3*sin(#2)}) circle (6pt);}{}
}

\newcommand\Poligon[4]{
    \StrCount{#1}{,}[\arrlength]
    \pgfmathsetmacro{\arrlength}{\arrlength + 1}
    \foreach[count=\i] \v in {#1} {
        \pgfmathsetmacro{\a}{(\i-1)*360/\arrlength + #2}
        \colorFillDraw{\v}{\a}{#3};
    }
  \ifthenelse{#4 > 0}{
    \foreach \phit in {1,...,\arrlength} {
        \pgfmathsetmacro{\phi}{(\phit-1)*360/\arrlength + #2}
        \pgfmathsetmacro{\phiP}{\phit*360/\arrlength + #2}
        \draw[black,line width=0.3mm] ({#3*cos(\phiP)}, {#3*sin(\phiP)}) -- ({#3*cos(\phi)}, {#3*sin(\phi)});
        \pgfmathsetmacro{\phiP}{\phi}
    }}{}
}

\newcommand\Conn[5][]{
    \draw[#1,line width=0.3mm] ({#2*cos(#3)}, {#2*sin(#3)}) -- ({#4*cos(#5)}, {#4*sin(#5)});
}

\title{Energy landscapes of small SK spin glasses}
\author{Imre Kondor$^{1,2}$, Gábor Papp$^3$}
\date{%
    {\small
    $^1$Complexity Science Hub, Vienna, Austria\\%
    $^2$London Mathematical Laboratory, London, UK\\%
    $^3$Institute for Physics and Astronomy, Eötvös Loránd University, Budapest, Hungary%
    }\\[2ex]%
    \today
}

\begin{document}
\maketitle

\begin{abstract}
We study the $\pm J$ SK model for small $N$'s up to $N=9$. We sort the $2^{N(N-1)/2}$ possible realizations of the coupling matrix into equivalence classes according to the gauge symmetry and the permutation symmetry and determine the energy spectra for each of these classes. We also study the energy landscape in these small systems and find that the elements of the hierarchical organization of ground states 
start to appear in some samples already for $N$'s as small as 6. 
\end{abstract}

\section{Introduction}

The Sherrington-Kirkpatrick model~\cite{SK}, the mean field model of spin glasses, has been the subject of countless studies over the past nearly 50 years. Its solution inspired tremendous progress in a large number of disciplines \cite{Far_beyond} and earned the 2021 Nobel prize to Giorgio Parisi. It may seem that nothing has been left to learn about the SK model by now.
This is not quite true. The overwhelming majority of works on the SK model, including its solution by Parisi~\cite{Parisi1979,Parisi1980} and the rigorous work by Guerra~\cite{Guerra}, Talagrand~\cite{Talagrand} and others, were all concerned with the thermodynamic limit, where the number $N$ of spins goes to infinity. In this asymptotic limit important simplifications take place, e.g. thermodynamic quantities self-average, that is become independent of the realization of disorder. This allows one to approach the problem by averaging over the randomness \cite{Brout1959}. Even numerical work on spin glasses, which is necessarily done on finite samples, is performed with quenched averaging and with the ultimate goal of extrapolating to $N\to\infty$ in mind.

However, in a large number of problems where the spin glass metaphor is relevant, the system size is definitely not macroscopic: there are no $10^{24}$ competing agents in an economy, or species fighting for survival in an ecosystem. 
In several contexts, it may be perfectly justified to study finite, or even small systems. (Think of the example of a family with positive or negative emotional links between the family members who, in addition, are strongly politically committed,  left or right, not necessarily overlapping with their emotional relationships. This is evidently a variant of the case of sorting the employees of a firm in two groups in the introduction of Mézard, Parisi and Virasoro's famous book~\cite{MPV}, but $N$ is naturally small in our example.) In such systems self-averaging cannot hold and quenched averaging cannot be justified; instead, the individual samples have to be studied one by one. 

When dealing with a random optimization problem, mathematics often seeks guarantee against the worst case, while statistical physics is focused on the typical, i.e. the average behavior. We argue that in some real life applications it may be relevant to study the concrete, particular case in question. Of course, the combinatorial explosion of the number of different cases sets a limit to the sizes one can reach in such an endeavor, which makes it important to exploit the symmetries and classify the different cases into classes having identical energy spectra, hopefully more manageable in number. We call these classes isospectral in the following. This classification is one of the main goals of this paper. In this task, we make use of some elementary concepts and results borrowed from the theory of signed graphs \cite{Harary1953, Zaslavsky1982}.

We have also determined the energies and their degree of degeneracy for each of these isospectral classes, as well as the corresponding spin configurations. These data allow us to construct the energy landscape at low energies, and also the local energy minima if they appear at higher energies. Because of the huge number of cases, these results are necessarily partial, but illustrate the  point sufficiently.

It is often stated that ordering or a phase transition can only take place in an infinitely large system. In practice, already fairly small systems anticipate some characteristic features of the ordered phase in an elementary form if the temperature is low enough. 
The special sort of order discovered by Parisi \cite{Parisi1979,Parisi1980} in the SK model \emph{in the large $N$ limit and on average} is complex to an earlier never seen degree. It is interesting to understand how such a complex structure begins to build up in a finite system. Given the messy structure of a disordered system, it is not surprising that, as pointed out by \cite{Ciliberti2004}, one has to go to very large sizes to observe the emergence of ultrametric geometry based on the metric of overlaps. However, the fragmentation of phase space and the elements of the hierarchical organization of ground states that we regard as the harbingers of the future ultrametric order start to appear already at $N=6$. In this paper, we display some examples of this phenomenon.

A related question is the concentration of measure \cite{Talagrand}, the fundamental reason behind self-averaging. We find that some invariance classes appear with much higher multiplicity than others when the system size is larger than $N=6$. For example, at $N=9$ the number of isospectral classes is 1625; out of these 17 appear with multiplicities well beyond a million. Nevertheless, we cannot speak of the dominance of these large-multiplicity classes, because their total multiplicity is still small compared with the total multiplicity $2^{(N-1)(N-2)/2} = 2^{28}=  268435456$ of all the classes - self-averaging is still very very far away.   

The plan of the paper is as follows. In Sec. 2 we set up the problem and fix notation. In Sec. 3 we make a few remarks and derive some sum rules that are valid for every $N$. In Sec. 4 we present the classification of the configurations and the energy spectra in each class up to $N=6$. Because the number of cases grows extremely fast, the results for $N=7, 8$, and 9 cannot fit into the paper, so we have set up a site to store this material~\cite{elte_storage_site}.

In Sec. 5 we exhibit several examples of the energy surface and show how the degenerate ground states divide the phase space into basins of attraction (very small, at these low $N$'s), and how they start to organize into families, the beginning of the evolution toward the ultrametric structure for large $N$'s. The organization of ground states is not the only feature that reflects the lack of convexity of the energy surface: we also find several local minima, saddle points and extended isolated clusters at higher energies. A striking example is a torus consisting of 60 alternating first and second excited states and surrounded by states of higher energy in one of the $N=8$ classes. 

Finally, Sec. 6 is a short summary. 


\section{Setting up the problem}
We study the $\pm J$ SK model, the simplest spin glass model placed on the complete graph. The spins will be standard Ising spins, $s_i=\pm 1$, $i=1,2,\ldots N$, while the couplings $J$ will also be considered simple binary variables, $J_{i,j}= \pm 1$, $i,j=1,2,\ldots,N$ for $i\ne j$ and $J_{i,i}=0$.

The properties we are going to calculate are just the zero temperature quantities, including the spectra of energies for the different realizations of the coupling matrix and the phase space map of the ground states and low lying excited states. 

For a given $N$ the number of couplings on the complete graph is $N(N-1)/2$, so the number of different realizations of signed samples is $2^{N(N-1)/2}$. For a given configuration, we will have $2^N$ spin states ($N$-vectors with components $\pm1$). These numbers grow extremely fast with $N$. In order to investigate the energy landscape, we have to store not only the energies but also the corresponding spin configurations. The volume of data quickly exceeds what can be surveyed, therefore we have to restrict this study to really small systems up to $N=9$. However, some of the results we find can be generalized to any $N$. 


The Hamiltonian of the system is the standard SK Hamiltonian:
\begin{equation}\label{Hamiltonian}
H = - \sum_{i<j} s_iJ_{i,j}s_j.
\end{equation}
The couplings $J_{i,j}=J_{j,i}$ and $J_{i,i}=0$ will be defined to take on the values $+1$ and $-1$ in every possible way, defining the $2^{N(N-1)/2}$ different configurations mentioned above. As we do not plan to play statistical mechanics or consider large systems here, we do not need to use the usual normalization ensuring the extensivity of the energy. This has the advantage of dealing with integers throughout.

The Hamiltonian is a functional of the coupling matrix $J_{i,j}$. For a given $J$ and depending on the spin vector $s$, it takes a set of values that we will refer to as the energy spectrum of $H$ (not to be mistaken for the spectrum of the matrix $J$). For a given $J$ the value of $H$ will be different according to the spin vector considered. These energies will, in general, be degenerate. We will call the energies $\epsilon_i$ and their degeneracy $\mu_i$.

$H$ possesses some obvious symmetries that help to reduce the number of different cases to be considered. The first such symmetry is the reflection of the spins $s_i$ to $-s_i$. The Hamiltonian is obviously invariant w.r.t. this transformation.

A reflection of the couplings $J_{i,j}$ to $-J_{i,j}$ results in the reflection of the whole spectrum $\epsilon_i$ to $-\epsilon_i$  with the degeneracy $\mu_i$ remaining the same.

It is well known that $H$ is also invariant w.r.t. a gauge transformation defined as follows: let the matrix of gauge transformation be a diagonal matrix $G$ with diagonal elements $g_i =\pm1$. Such a matrix is clearly orthogonal. Multiply the spin vectors $s$ with this matrix to obtain $Gs$ and multiply the coupling matrix as $GJG^{-1}$. The value of the energies and degeneracies will be left invariant under this gauge transformation. 

The interaction matrix $J_{i,j}$ as defined above is the adjacency matrix of a signed complete graph. (We are dealing with complete graphs all through this paper.) Signed graphs were introduced in a social psychology context by Harary \cite{Harary1953}, but the same concepts were discussed earlier under a different terminology in Dénes König's book~\cite{Konig1937}.

The gauge transformation is the same as what is called the switching transformation in the field of signed graphs \cite{Harary1953, Zaslavsky1982}. In addition to the invariance of the Hamiltonian within a gauge invariant class, the spectra of interaction matrices themselves are also invariant. 

Finally, the problem is invariant w.r.t. the permutation of the indices $i$.

Of these symmetries the reflection is trivial. 

The gauge symmetry sorts the different cases into gauge classes with $2^{(N-1)}$ members in each. This allows us to group these equivalent configurations into a single gauge class and  deal with only one representative of this class in the following. This representative can be selected by fixing the sign on a spanning tree in the graph whose adjacency matrix is $J_{i,j}$~\cite{Burda_Heider}, for example by setting all the non-diagonal elements in the first row of the matrix to be 1. Sometimes it is expedient to fix the gauge this way, sometimes the original gauge symmetric form is more amenable for general considerations. An economic way of selecting the representative configuration is to reduce the number of negative couplings to its smallest value via a series of gauge transformations. This minimal signature is not unique, permutations of vertices can take it over into other equivalent configurations. Nevertheless, we will present one of these minimal signature graphs as the representative of the gauge equivalent class unless some symmetry considerations make it advantageous to display a different member of the class. We have determined these minimal signatures in all the cases investigated.

Grouping the gauge equivalent configurations into gauge classes reduces the number $2^{N(N-1)/2}$ of different configurations by a factor $2^{(N-1)}$, leaving $2^{(N-1)(N-2)/2}$ cases to study. 

Finally, the permutation of the vertices collects a number of different $J_{i,j}$’s into equivalence classes. Different gauge- and/or permutation-equivalent coupling configurations have the same energy spectra, but the reverse is not necessarily true. When we speak about equivalence classes in the following we always mean isospectral classes, i.e. sets of configurations with the same energy spectra. The number of different coupling configurations belonging to such a class will be called the multiplicity of the given class. The determination of these multiplicities can be a not entirely trivial combinatorial problem already for moderate $N$'s.

We have also determined the number of frustrated 3~-,~4~-, \ldots $N$-cycles for all classes up to $N=9$. Remarkably, we found that among the nearly 2000 graph classes surveyed there was only a single pair at $N=8$ that had the same number of all frustrated $k$-cycles, $k=3,\ldots, N$,  but belonged to different invariance classes. Otherwise the number of frustrated cycles was a useful tool to sort the equivalent graphs into invariance classes. The number of cycles will be displayed in the following up to $N=6$ in the main text and for $N=7,8,9$ on the web page~\cite{elte_storage_site}.

The classification of the configurations into equivalence classes according to the gauge symmetry and the permutation symmetry is parallel to similar efforts in signed graphs. We have not found such results for signed graphs except for $N=6$ \cite{Sehrawat_and_Bhattacharjya}. We go up to $N=9$ in this paper. 

Moreover, we have determined the energy spectra and the corresponding spin configurations in all these cases, which are of no concern for workers in the field of signed graphs. 

An old paper having some overlap with our results is due to Caliri et al.~\cite{Caliri1985}, whose main interest is in the thermodynamics of the system and the zeros of the partition function on the complex plane. In their Table 1 they list the weights of configurations for $N=5$. Our results for $N=5$ are in agreement with their classification and with their weights that correspond to what we call the multiplicities of these configurations. 

Boettcher and Kott~\cite{Boettcher2005} consider all the instances of the $\pm J$ SK model and also go up to $N=9$, as we do, but their focus is on the number of frustrated bonds and the distribution of ground state energies, different from ours.


\section{Some results for general \texorpdfstring{$N$}{N}}

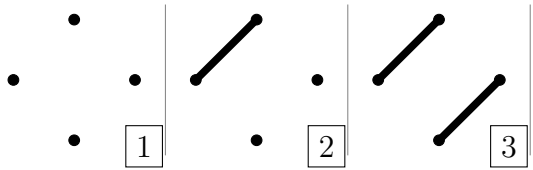
\begin{figure}[h]
    \centering
\begin{tikzpicture}[scale=0.4]
\setAllDots[black]{3}{4}{0}

\ConnN[black]{2}{4}{0}{2}{3}

\ConnN[black]{3}{4}{0}{1}{4}
\ConnN[black]{3}{4}{0}{2}{3}
\end{tikzpicture}
    \caption{Classes for $N=4$}
    \label{fig:N=4_configurations}
\end{figure}

We can illustrate some results valid for general $N$ on the example of the simple case of $N=4$ where there are only three independent equivalence classes. The corresponding interaction graphs are shown in Fig.~\ref{fig:N=4_configurations}. For higher $N$'s a complete graph can be very crowded, so we decided to adopt the convention that only the negative edges will be shown, the positive ones left understood. So the first graph in Fig.~\ref{fig:N=4_configurations} is in fact the complete graph with all positive edges (the "ferromagnet"), and it also represents all 8 of its gauge equivalent versions. 
Similarly, the second and third graph in Fig.~\ref{fig:N=4_configurations} are representatives of their 8-member gauge classes. In particular, the third one is gauge equivalent to the purely antiferromagnetic graph, with all its edges negative. Note that we display the representatives of these three classes with the minimal number of negative couplings, as promised. So the minimal signature in these three classes has zero, one, resp. two negative edges. 

The multiplicities shown in Table~\ref{tab:Table1} (and wherever multiplicities are mentioned later) are the numbers of permutation equivalent configurations remaining after factoring out the number of gauge equivalent configurations. The three cases left after collecting the gauge symmetric configurations into groups are shown in the following Table~\ref{tab:Table1}.

\begin{table}[h]
\centering\begin{tabular}{|c|c|c|c|c|}
\hline
Class&1&2&3\\
\hline
Multiplicity&1&6&1\\
\hline
No. of 3-cycles &4+, 0-&2+, 2-&0+, 4-\\
\hline
No. of 4-cycles &3+, 0-&1+, 2-&3+, 0-\\
\hline
6& & &6(2)\\
\hline
5& &4(2)&\\
\hline
4&2(6)&2(4)&\\
\hline
3&0(8)&0(4)&0(8)\\
\hline
2& &-2(4)&-2(6)\\
\hline
1& &-4(2)&\\
\hline
0&-6(2)& &\\
\hline
\end{tabular}

 \caption{\label{tab:Table1} Energies for $N=4$. In the third row we give the number of positive (balanced or not frustrated) triangles and the number of negative (frustrated) triangles in the graph of interactions. 
 For example, the entry 2+, 2- in the third row and third column means that there are two balanced and two frustrated triangles in a Class 2 graph. In the fourth row the same is given for the quadrangles. In the last seven rows in the first column the labels of the energy levels is given: 0 means ground state, 1 first excited state, etc.}
 \end{table}

The total number of cases for $N=4$ is $2^{N(N-1)/2}=2^6=64$. As we consider only one representative of each gauge class, we divide this by $2^{(N-1)}=8$. So we are left with 8 configurations to distribute among the permutation classes. It is easy to see that we have one purely ferromagnetic and one purely antiferromagnetic configuration, plus six permutation equivalent configurations, corresponding to the number of ways a negative bond can be placed on an $N=4$ complete graph (that is on the edges of a tetrahedron). 

These three classes are shown in the first row of Table~\ref{tab:Table1}. In the second row we exhibit the multiplicities 1, 6, and 1 of these classes. The third row shows the number of positive (or balanced, in the parlance of signed graph theory) and negative (frustrated) triangles (3-cycles). (In a positive triangle the product of signs on its edges is positive, in a negative triangle negative.) 
In the third row the numbers of positive and negative 4-cycles are given. We can see that for $N=4$ the number of negative triangles is sufficient to distinguish between the three classes.

The figures in the first column label the energy levels in this small system, the numbering going from 0 to 6. For general $N$ the possible energy values range from $-\frac{1}{2}N(N-1)$, the ground state energy of the purely ferromagnetic model, to $\frac{1}{2}N(N-1)$, the maximal energy of the pure antiferromagnet. Since the energy steps are 2 units, this makes $\frac{1}{2}N(N-1) + 1$ possible energy levels. Note that not each of these energy levels are necessarily occupied in a given class. 

The second column shows the energy spectrum in Class 1 with the degeneracy of the energy shown in parentheses. So the ground state energy in Class 1 is $-~6$, two-fold degenerate, the energy of the first excited state is zero with a degeneracy of 8, while the highest energy is 2, six times degenerate. In the third column the same data are listed for Class 2, in the fourth the same for Class 3.

For general $N$ the ferromagnetic spectrum goes from $-\frac{1}{2}N(N-1)$ to $\frac{1}{2}N$ for $N$ even, and to $\frac{1}{2}(N-1)$ for $N$ odd. The antiferromagnetic spectrum can be obtained from this by reflection with respect to zero.  The spectrum of the other classes falls between these two extremes.

\subsection{Vertical sum rules}
The sum of the degeneracies in each column in Table~\ref{tab:Table1} ($6+8+2$ in Class 1, $2+4+4+4+2$ in Class 2, and $2+8+6$ in Class 3) is 16; for general $N$ it is $2^N$, the dimension of the phase space of $N$ binary spin variables. This obviously holds for any configuration.

If we multiply the energies with their degeneracies in any class (for example $2\times6$ + $0\times8$ - $6\times2$ in Class 1), we get zero. This is true in general: the sum of the products of the energies and multiplicities is zero for each configuration, for any $N$. This is explained by the properties of the spin vectors and by the coupling matrix $J_{i,j}$ being zero along its diagonal. Similarly, one can determine the sum of the product of the degeneracies with the squared energies $(2\times9+6\times1=24)$; for generic $N$ the result works out to be $2^N\times N(N-1)/2= 2^{N-1}$\rm{Tr}$J^2$. The sum of the degeneracies times the energies cubed is $-2^N$\rm{Tr}$J^3$. One can derive similar sum rules for any higher power of the energies, but they quickly become very complicated expressions constructed from the algebraic invariants of the permutation group.

\subsection{Horizontal sum rule}

The sum of the multiplicities in the second row of Table~\ref{tab:Table1} is 8; for general $N$ it is $2^{(N-1)(N-2)/2}$.

Let us now determine the number of times a given energy appears in the whole ensemble of configurations. This is obtained by multiplying the multiplicity of the configuration by the degeneracy of the given energy and summing over the configurations. The energy -~6 at level zero appears 2 times. The energy -~4 at level one $6\times2$ times, the energy -~2 at level two $6\times4+1\times6$ times, etc.. Thus the energy at level $i$ is seen to show up $2\times \binom{6}{i}$ times. For generic $N$ there are a large number of configurations, all with their rather haphazard looking multiplicities and the different energies with different degeneracies, but the product of multiplicities and degeneracies at a given level $i$ of energy summed across the configurations will be given by the simple formula 
$2\binom{N(N-1)/2}{i}$. (The prefactor would be $2^N$ if we had not divided the number of configurations by the multiplicity of the gauge classes.)

These sum rules provide checks on the energy spectra that become fairly complicated as $N$ increases.

\section{The energy spectra}
\label{sec:en_spectra}
Apart from the smallest $N$'s, making sure that we generate all the configurations and count their multiplicities correctly is not completely trivial. It is expedient to generate them by a systematic method. The one we used, applying a code kindly provided by Matteo Marsili, is as follows: First, we fix the gauge by setting the first line of the interaction matrix as $0,1,1,\dots,1$.  This reduces the number of configurations by a factor of $2^{N-1}$, as mentioned earlier. The upper triangular part of the remaining submatrix is then produced as we count up to $2^{(N-1)(N-2)/2}$ considering the binary representation of the number (counter) as the row-major order of the triangular matrix, associating a positive bond with, say 0 and the negative one with 1. This procedure generates all the different configurations. Several of these will be equivalent (will have the same spectrum) because of the permutation symmetry, and we have to collect them into equivalence classes. The number of configurations in such a class is the multiplicity of the class. In the following, we will explicitly deal with only one member of the class, the other members being accounted for by the multiplicity.

Unfortunately, generating the configurations as suggested above results in a somewhat random order of their appearance. The first configuration will always be the purely ferromagnetic $J$ (all elements +1), and the last one, the pure antiferromagnetic configuration with all its elements equal to $-1$ (or one of its gauge transforms with less negative matrix elements), but apart from these two extremes the others come out in a more or less random fashion. It is desirable to arrange the different configurations in some logical order. The first idea that may come to mind is to order the configurations according to the number of negative bonds. However, because of the gauge symmetry the number of negative bonds do not provide a consistent classification: graphs containing a different number of negative bonds often belong to the same equivalence class. (For example, we have seen in Fig.~\ref{fig:N=4_configurations} that for $N=4$ a graph with merely two negative edges can stand for one with all six edges being negative.) We found that the lexicographic order according to the number of negative triangles, quadrangles, pentagons, etc. in the complete graph of the interaction matrix provides a possible ordering principle together with an insight into the geometry of interactions.  

Let us now suppose we have a given interaction matrix and want to calculate its spectrum. In order to do this, we have to apply the formula~\eqref{Hamiltonian} for each of the $2^N$ different spin vectors. If we form an $N\times2^N$ matrix $S$ out of these spin vectors and call the interaction matrix $J$, the spectrum will be given by the diagonal elements of -$\frac{1}{2}S^TJS$. (The superscript $T$ means transpose.)

For $N=4$ we have already seen the results in Table~\ref{tab:Table1}. Now we turn to the case of $N=5$.

\subsection{Invariance classes and energy spectra for \texorpdfstring{$N$}{N}=5}

\begin{figure}[h]
    \centering
\begin{tikzpicture}[scale=0.4]
\setAllDots[black]{7}{5}{18}
\ConnN[black]{2}{5}{18}{4}{5}
\ConnN[black]{3}{5}{18}{4}{5}
\ConnN[black]{3}{5}{18}{5}{1}
\ConnN[black]{4}{5}{18}{4}{5}
\ConnN[black]{4}{5}{18}{5}{1}
\ConnN[black]{4}{5}{18}{1}{2}
\ConnN[black]{5}{5}{18}{4}{5}
\ConnN[black]{5}{5}{18}{1}{2}
\ConnN[black]{6}{5}{18}{4}{5}
\ConnN[black]{6}{5}{18}{5}{1}
\ConnN[black]{6}{5}{18}{2}{3}
\ConnN[black]{7}{5}{18}{4}{5}
\ConnN[black]{7}{5}{18}{5}{1}
\ConnN[black]{7}{5}{18}{1}{4}
\ConnN[black]{7}{5}{18}{2}{3}
\end{tikzpicture}
    \caption{Classes for $N=5$.}
    \label{fig:N=5_configurations}
\end{figure}
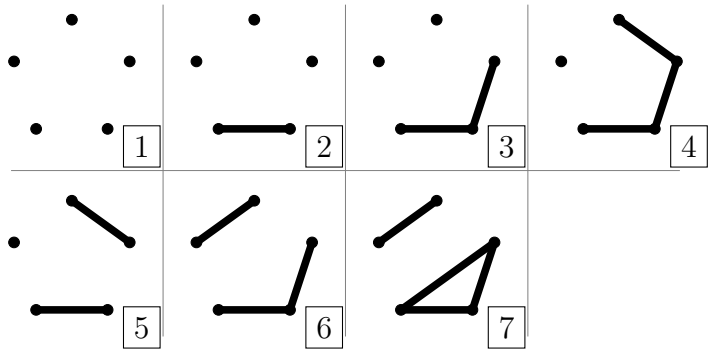

The seven $N=5$ configurations are exhibited in Fig.~\ref{fig:N=5_configurations}, their  energies and multiplicities are displayed in Table~\ref{tab:Table2}.

\begin{table}[h]
\centering\begin{tabular}{|c|c|c|c|c|c|c|c|}
\hline
Class&1&2&3&4&5&6&7\\
\hline
Multiplicity&1&10&15&12&15&10&1\\
\hline
No. of 3-cycles&10+, 0-&7+, 3-&6+, 4-&5+, 5-&4+, 6-&3+, 7-&0+, 10-\\
\hline
No. of 4-cycles&15+, 0-&9+, 6-&7+, 8-&5+, 10-&7+, 8-&9+, 6-&15+, 0-\\
\hline
No. of 5-cycles&12+, 0-&6+, 6-&4+, 8-&6+, 6-&8+, 4-&6+, 6-&0+, 12-\\
\hline
10&&&&&&&10(2)\\
\hline
9&&&&&&8(2)&\\
\hline
8&&&6(2)&&6(4)&&\\
\hline
7&&4(8)&&4(10)&&4(4)&\\
\hline
6&2(20)&&2(16)&&2(10)&&2(10)\\
\hline
5&&0(18)&&0(12)&&0(18)&\\
\hline
4&-2(10)&&-2(10)&&-2(16)&&-2(20)\\
\hline
3&&-4(4)&&-4(10)&&-4(8)&\\
\hline
2&&&-6(4)&&-6(2)&&\\
\hline
1&&-8(2)&&&&&\\
\hline
0&-10(2)&&&&&&\\
\hline
\end{tabular}
 
 \caption{\label{tab:Table2} Multiplicities, number of balanced and frustrated 3-, 4-, and 5- cycles, and energies for $N=5$.} 
 \end{table}

Note that at $N=5$ the number of frustrated triangles is still sufficient to fully differentiate between the symmetry classes. This will not be the case for higher $N$'s, as shown in Table~\ref{tab:Table3} in the next subsection.

\subsection{Invariance classes and energy spectra for \texorpdfstring{$N$}{N}=6}
The number of invariance classes for $N=6$ is 16 . Their multiplicities, the number of frustrated 3-, 4-, and 5-cycles are displayed in the first 5 rows in Table~\ref{tab:Table3}. Note that for lack of space only the number of frustrated cycles is given. 

A little reflection shows that the total number of $k$-cycles in a complete graph on $N$ vertices is
\begin{equation}\label{number of cycles}
C(N,k)=\frac{1}{2k}\frac{N!}{(N-k)!} \,.
\end{equation}

Hence the total number of 3, 4, resp. 5 cycles in a six-vertex complete graph is 21, 45, resp. 72. The number of balanced cycles can be calculated by subtracting the number of frustrated cycles from the total.

For $N=6$ the number of frustrated triangles is already not enough to distinguish the different invariance classes, because, for example, in classes 4, 5 and 6 this number is the same 8, so we have to determine the number of longer cycles to see the difference between them. The number of quadrangles is still not enough, because it coincides between classes 5 and 6. We also have to count the pentagons in order to resolve these coincidences. 

Unfortunately, we have not found any formula for the number of positive and negative $k$-cycles for general $N$ beyond the case of the triangles where the difference between the number of positive and negative triangles is given by $\frac{1}{6}Tr J^3$. So one has to determine the number of negative cycles by direct counting, which we have performed up to $N=9$. As $N$ increases further, this task quickly becomes prohibitive.

In the paper by Sehrawat and Bhattacharjya~\cite{Sehrawat_and_Bhattacharjya} the classes of the $N=6$ complete graphs were identified precisely on the basis of frustrated cycles. The numbers shown in rows 3, 4 and 5 in Table~\ref{tab:Table3} are quoted from their paper (with a typo corrected). The order in which these classes are displayed in Table~\ref{tab:Table3} is, however, different from theirs, because we wanted to arrange the classes in lexicographic order according to the number of frustrated cycles.

The rest of the table shows the energy spectra of these classes, which were of course not included in the signed graph analysis in \cite{Sehrawat_and_Bhattacharjya}. In the first column the energy values are listed and the numbers in the rows show the degree of degeneracy of these energies in the various classes. As mentioned earlier, these spectra are also invariants of the classes.

\begin{figure}[h]
    \centering
\begin{tikzpicture}[scale=0.4]
\setAllDots[black]{16}{6}{30}

\ConnN[black]{2}{6}{30}{1}{6}

\ConnN[black]{3}{6}{30}{1}{6}
\ConnN[black]{3}{6}{30}{1}{2}

\ConnN[black]{4}{6}{30}{1}{4}
\ConnN[black]{4}{6}{30}{2}{3}

\ConnN[black]{5}{6}{30}{6}{1}
\ConnN[black]{5}{6}{30}{1}{2}
\ConnN[black]{5}{6}{30}{2}{3}

\ConnN[black]{6}{6}{30}{6}{1}
\ConnN[black]{6}{6}{30}{1}{2}
\ConnN[black]{6}{6}{30}{2}{3}
\ConnN[black]{6}{6}{30}{3}{6}

\ConnN[black]{7}{6}{30}{6}{1}
\ConnN[black]{7}{6}{30}{1}{2}
\ConnN[black]{7}{6}{30}{2}{6}

\ConnN[black]{8}{6}{30}{1}{2}
\ConnN[black]{8}{6}{30}{2}{3}
\ConnN[black]{8}{6}{30}{4}{6}

\ConnN[black]{9}{6}{30}{6}{1}
\ConnN[black]{9}{6}{30}{1}{2}
\ConnN[black]{9}{6}{30}{2}{3}
\ConnN[black]{9}{6}{30}{3}{4}

\ConnN[black]{10}{6}{30}{6}{1}
\ConnN[black]{10}{6}{30}{1}{2}
\ConnN[black]{10}{6}{30}{2}{3}
\ConnN[black]{10}{6}{30}{3}{4}
\ConnN[black]{10}{6}{30}{4}{6}

\ConnN[black]{11}{6}{30}{1}{2}
\ConnN[black]{11}{6}{30}{2}{3}
\ConnN[black]{11}{6}{30}{4}{5}
\ConnN[black]{11}{6}{30}{5}{6}

\ConnN[black]{12}{6}{30}{1}{4}
\ConnN[black]{12}{6}{30}{2}{3}
\ConnN[black]{12}{6}{30}{5}{6}


\ConnN[black]{13}{6}{30}{5}{6}
\ConnN[black]{13}{6}{30}{6}{1}
\ConnN[black]{13}{6}{30}{1}{2}
\ConnN[black]{13}{6}{30}{3}{4}

\ConnN[black]{14}{6}{30}{1}{2}
\ConnN[black]{14}{6}{30}{2}{3}
\ConnN[black]{14}{6}{30}{3}{1}
\ConnN[black]{14}{6}{30}{4}{6}

\ConnN[black]{15}{6}{30}{1}{2}
\ConnN[black]{15}{6}{30}{2}{6}
\ConnN[black]{15}{6}{30}{6}{1}
\ConnN[black]{15}{6}{30}{3}{4}
\ConnN[black]{15}{6}{30}{4}{5}

\ConnN[black]{16}{6}{30}{1}{2}
\ConnN[black]{16}{6}{30}{2}{6}
\ConnN[black]{16}{6}{30}{6}{1}
\ConnN[black]{16}{6}{30}{3}{4}
\ConnN[black]{16}{6}{30}{4}{5}
\ConnN[black]{16}{6}{30}{5}{3}

\end{tikzpicture}
    \caption{Classes for $N=6$.}
    \label{fig:N=6_configurations}
\end{figure}
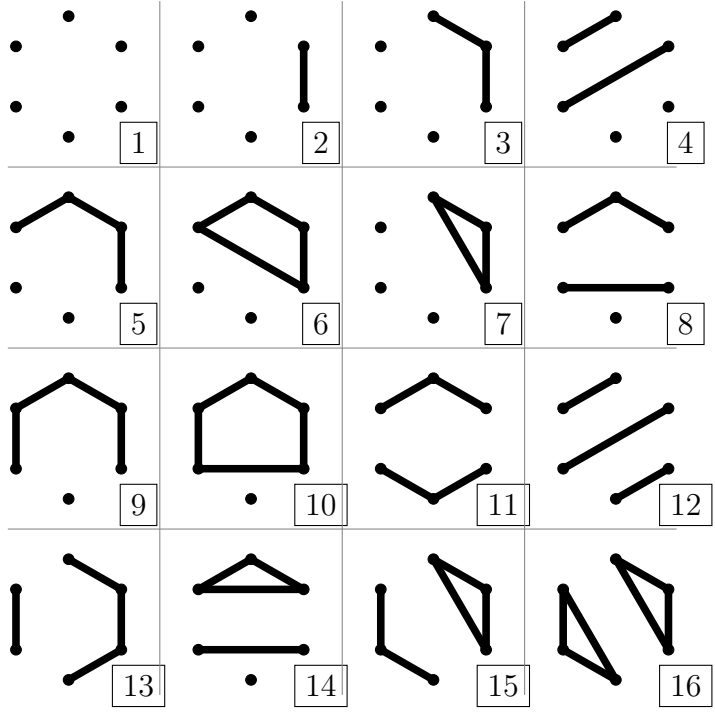

\begin{table}[h]
\scriptsize
\centering\begin{tabular}{|c|c|c|c|c|c|c|c|c|c|c|c|c|c|c|c|c|}
\hline
Class&1&2&3&4&5&6&7&8&9&10&11&12&13&14&15&16\\
\hline
Multiplicity&1&15&60&45&180&15&20&180&180&12&45&15&180&60&15&1\\
\hline
Frustrated 3-cycles&0&4&6&8&8&8&10&10&10&10&12&12&12&14&16&20\\
\hline
Frustrated 4-cycles&0&12&18&20&24&24&18&22&26&30&20&24&24&18&12&0\\
\hline
Frustrated 5-cycles&0&24&36${}^*$&32&40&48&36&36&36&36&40&24&32&36&48&72\\
\hline
Frustrated 6-cycles&0&24&36&24&32&32&36&28&28&20&24&32&32&36&24&0\\
\hline
Energy&&&&&&&&&&&&&&&&\\
\hline
15&&&&&&&&&&&&&&&&2\\
\hline
13&&&&&&&&&&&&&&&2&\\
\hline
11&&&&&&&&&&&2&&&2&&\\
\hline
9&&&&&&2&2&2&&&&&2&2&&\\
\hline
7&&&2&4&2&&6&2&4&&&6&4&&4&\\
\hline
5&&8&6&6&8&&&4&8&12&8&12&4&4&&12\\
\hline
3&20&14&12&8&12&20&&12&8&20&8&&8&8&8&\\
\hline
1&30&12&16&16&12&24&24&12&12&&12&&12&12&16&\\
\hline
-1&&16&12&12&12&&24&12&12&&16&24&12&16&12&30\\
\hline
-3&&8&8&8&8&&&12&8&20&8&20&12&12&14&20\\
\hline
-5&12&&4&8&4&12&&4&8&12&6&&8&6&8&\\
\hline
-7&&4&&&4&6&6&2&4&&4&&2&2&&\\
\hline
-9&&&2&&2&&2&2&&&&2&&&&\\
\hline
-11&&&2&2&&&&&&&&&&&&\\
\hline
-13&&2&&&&&&&&&&&&&&\\
\hline
-15&2&&&&&&&&&&&&&&&\\
\hline
\end{tabular} 

 \caption{\label{tab:Table3} Multiplicities, frustrated 3-, 4-, 5-, and 6-cycles and energies for $N=6$. For lack of space, only the energies are listed in the first column, while their degeneracies are shown in the Table. For example, the degeneracy of energy 3 in Class 1 is 20, in Class 2 it is 14, in Class 3 it is 12, etc. ${}^*$The number of frustrated pentagons in Class 3 is 36, at variance with the result in~\cite{Sehrawat_and_Bhattacharjya} who give 24. We checked their results and found general agreement except at this particular point, which may be a typo in their paper. The value of 36 is also dictated by symmetry: the interaction matrix in Class 3 is the negative of that in Class 14. We have also added the results for the number of 6-cycles, which were not included in ~\cite{Sehrawat_and_Bhattacharjya}}.
\end{table}

 We have determined the isospectral classes for $N=7,~8$ and $9$ as well. Their numbers are 54, 219, and 1625, respectively. We also have the number of frustrated cycles and the energy spectra for these cases.  
 Such a large amount of data cannot be squeezed into this paper, therefore we placed it on the web site~\cite{elte_storage_site}.

 \section {Energy maps}

 In this Section we present a sample of energy maps for different $N$'s. In Subsection 5.1 we focus on the ground states and adjacent excited states to illustrate how the degenerate ground states begin to organize, already for small $N$'s, into clusters anticipating some elementary features of the ultrametric structure. In view of the large number of cases we will not go into details in each class, only in those that present some interesting properties.
 In Subsection 5.2 we make a few comments on higher energy minima and other isolated clusters.

 \subsection{Maps of ground states and low lying states}

 \subsubsection{The case of \texorpdfstring{$N$}{N}~=~3}
\begin{figure}[h]
    \centering
\begin{tikzpicture}[scale=0.45]
\useasboundingbox (-3,-3) rectangle (6,4);
\Poligon{1,1,0,1}{0}{3}{0}
\Conn[black]{3}{180}{3}{90}
\Conn[black]{3}{90}{3}{0}
\Conn[black]{3}{0}{3}{270}
\Conn[black]{3}{270}{3}{180}
\node[draw] at (-3,-3) {a};
\begin{scope}[shift={(3,0)}]
  \Poligon{0,1,1,1}{0}{3}{0}
  \Conn[black]{3}{180}{3}{90}
  \Conn[black]{3}{90}{3}{0}
  \Conn[black]{3}{0}{3}{270}
  \Conn[black]{3}{270}{3}{180}
\end{scope}
\draw[black,line width=0.3mm] (-3,0) -- (0,0);
\draw[black,line width=0.3mm] (3,0) -- (6,0);
\draw[black,line width=0.3mm] (0,3) -- (3,3);
\draw[black,line width=0.3mm] (0,-3) -- (3,-3);
\end{tikzpicture}
\hspace*{1cm}
\begin{tikzpicture}[scale=0.45]
\useasboundingbox (-3,-3) rectangle (6,4);
\Poligon{0,0,1,0}{0}{3}{0}
\Conn[black]{3}{180}{3}{90}
\Conn[black]{3}{90}{3}{0}
\Conn[black]{3}{0}{3}{270}
\Conn[black]{3}{270}{3}{180}
\node[draw] at (-3,-3) {b};
\begin{scope}[shift={(3,0)}]
  \Poligon{1,0,0,0}{0}{3}{0}
  \Conn[black]{3}{180}{3}{90}
  \Conn[black]{3}{90}{3}{0}
  \Conn[black]{3}{0}{3}{270}
  \Conn[black]{3}{270}{3}{180}
\end{scope}
\draw[black,line width=0.3mm] (-3,0) -- (0,0);
\draw[black,line width=0.3mm] (3,0) -- (6,0);
\draw[black,line width=0.3mm] (0,3) -- (3,3);
\draw[black,line width=0.3mm] (0,-3) -- (3,-3);
\end{tikzpicture} 
    \caption{Phase space maps for $N=3$ (black: ground state, blue: 1st excited state), (a) the ferromagnet, (b) the antiferromagnet. The dots represent the spin states, the links connect states that are next neighbors in phase space, which means they differ in the flip of a single spin.}
    \label{fig:N=3_space_map}
\end{figure}
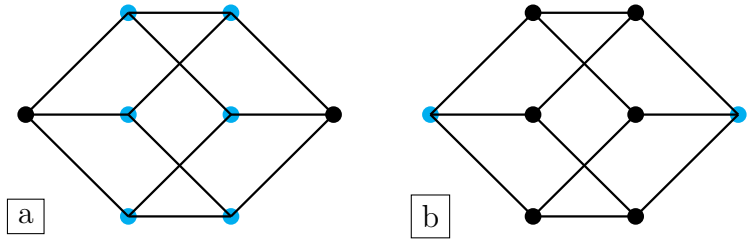

 Let us begin with the simplest possible case with $N=3$. There are just two classes here: the one with all couplings positive (a pure ferromagnet) and its gauge transformed variants, and the one with all couplings negative (the pure antiferromagnet) again with its gauge transforms. It is sufficient to consider a single representative of each class. For $N=3$ the phase space is just the ordinary cube that is easy to draw with all 8 spin states shown as in Fig.~\ref{fig:N=3_space_map}. The vertices in this graph correspond to the spin states; the edges connect neighboring spin states that differ only in the sign of a single spin. 

 The figure on the left in Fig.~\ref{fig:N=3_space_map} shows the phase space map of the $N=3$ ferromagnet. There are two degenerate ground states of energy $\epsilon_0~=-3$ and a six-fold degenerate excited state with energy $\epsilon_1~=+1$. The figure on the right in Fig.~\ref{fig:N=3_space_map} is the map of the states of the antiferromagnet. This map is the reverse of that of the ferromagnet: there is a six-fold degenerate ground state ($\epsilon_0~=-1$) and a two-fold degenerate excited state ($\epsilon_1~=+3$). The antiferromagnetic ground states are all connected with each other in a network. This is a special case of the general property of antiferromagnetic ground states for odd $N$: as can easily be seen, the ground state energy is $\epsilon_0~=~-\frac{1}{2}(N-1)$ and the degeneracy of the ground state is $\mu_0~=~\binom{N}{\frac{N+1}{2}} + \binom{N}{\frac{N-1}{2}}$. These ground states are connected to each other by $\frac{N+1}{2}$ links. The remaining $\frac{N-1}{2}$ couplings connect the ground states to the first excited states.

 The ferromagnetic case is just the opposite of the antiferromagnetic one, it is the highest excited states here that are the most numerous and connected with each other by  $\frac{N+1}{2}$ links, with an occasional path leading down toward the ground state. This arrangement can be called a plateau with some escape routes. If we think of an optimization task with such a phase space and initiate the search on the plateau, our program may spend a long time wandering around at the top before finding a way to descend.

 \subsubsection{The case of \texorpdfstring{$N$}{N}~=~4}

\begin{figure}[h]
     \centering
\begin{tikzpicture}[scale=0.6]
\useasboundingbox (-3,-3) rectangle (3,4);
\Poligon{0,1,2,1,0,1,2,1}{0}{3}{1}
\Poligon{2,1,2,1,2,1,2,1}{0}{1.5}{0}
\foreach \phi in {0,90,180,270} {
    \Conn[black]{3}{\phi}{1.5}{\phi+45}
    \Conn[black]{1.5}{\phi+45}{3}{\phi+90}
}
\foreach \phi in {0,90,180,270} {
    \Conn[black]{3}{\phi+45}{1.5}{\phi+90}
    \Conn[black]{1.5}{\phi+90}{3}{\phi+135}
}
\foreach \phi in {0,45,90,135,180,225,270,315} {
    \Conn[black]{1.5}{\phi}{1.5}{\phi+3*45}
}
\node[draw] at (-3,-3) {a};
\end{tikzpicture}
\hspace*{6mm}
\begin{tikzpicture}[scale=0.6]
\useasboundingbox (-3,-3) rectangle (3,4);
\Poligon{0,1,2,1,0,1,2,1}{0}{3}{1}
\Poligon{4,3,2,3,4,3,2,3}{0}{1.5}{0}
\foreach \phi in {0,90,180,270} {
    \Conn[black]{3}{\phi}{1.5}{\phi+45}
    \Conn[black]{1.5}{\phi+45}{3}{\phi+90}
}
\foreach \phi in {0,90,180,270} {
    \Conn[black]{3}{\phi+45}{1.5}{\phi+90}
    \Conn[black]{1.5}{\phi+90}{3}{\phi+135}
}
\foreach \phi in {0,45,90,135,180,225,270,315} {
    \Conn[black]{1.5}{\phi}{1.5}{\phi+3*45}
}
\node[draw] at (-3,-3) {b};
\end{tikzpicture}
\hspace*{6mm}
\begin{tikzpicture}[scale=0.6]
\useasboundingbox (-3,-3) rectangle (3,4);
\Poligon{2,1,0,1,2,1,0,1}{0}{3}{1}
\Poligon{0,1,0,1,0,1,0,1}{0}{1.5}{0}
\foreach \phi in {0,90,180,270} {
    \Conn[black]{3}{\phi}{1.5}{\phi+45}
    \Conn[black]{1.5}{\phi+45}{3}{\phi+90}
}
\foreach \phi in {0,90,180,270} {
    \Conn[black]{3}{\phi+45}{1.5}{\phi+90}
    \Conn[black]{1.5}{\phi+90}{3}{\phi+135}
}
\foreach \phi in {0,45,90,135,180,225,270,315} {
    \Conn[black]{1.5}{\phi}{1.5}{\phi+3*45}
}
\node[draw] at (-3,-3) {c};
\end{tikzpicture}
     \caption{Phase space maps for $N=4$ (black: ground state, blue: 1st excited state, green: 2nd excited state, yellow: 3rd excited state, orange: 4th excited state).}
     \label{fig:N=4_space_map}
 \end{figure}
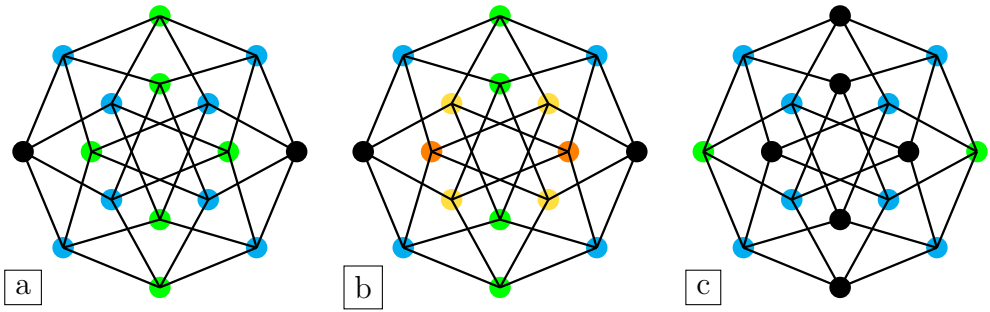

 There are three configurations for $N=4$. Their phase space map can still be shown in full (including all the 16 states) for this small $N$, as in Fig.~\ref{fig:N=4_space_map}. 

 The left Fig.~\ref{fig:N=4_space_map}a shows the $N=4$ ferromagnet, Class 1 in Table~\ref{tab:Table1}. This map corresponds to our usual way of speaking about an Ising ferromagnet: the paths starting from one of the ground states lead through the highest energy states before arriving at the "other half of phase space" and eventually at the other ground state. (The meaning of the two halves of phase space is obvious: spin vectors whose components sum to a positive value belong to one half of the phase space, those with a negative sum of their components to the other. For even $N$ there are vectors whose components sum to zero; these lie in the transition zone between the two halves.)

 Fig.~\ref{fig:N=4_space_map}b shows the effect of changing the sign of one coupling to negative. This slightly modifies the energy map, see Class 2 in Table~\ref{tab:Table1}: the ground state energy is increased, the paths leading to the other ground state can avoid the highest energy states and the energy gap to climb is diminished, 
 but the clear distinction between the two halves of phase space still remains.

 This state of affairs changes when we consider the fully frustrated configuration whose map is given in Fig.~\ref{fig:N=4_space_map}c. The ground states are now six-fold degenerate (Class 3 in Table~\ref{tab:Table1}) and can be reached from each other by going through the first excited states, so the communication between the regions of phase space is much easier and the distinction between its two halves is nearly meaningless. (Of course, this is true only as long as there is no external field present, but that is beyond the scope of the present paper.)

 This leads us to an easy generalization again. For a general even $N$ the ground state energy of the pure antiferromagnetic configuration (and its gauge and permutation equivalents) is $\epsilon_0~=~-N/2$ and its degeneracy is $\mu_0~=~\binom{N}{\frac{N}{2}}$. Every ground state has $N$ next neighbors among the first excited states and does not have other neighbors apart from those.

 Conversely, the highest excited states in the purely ferromagnetic configuration are connected to $N$ states with an energy one step below. So, in contrast to the case of an odd $N$, a search initiated at the highest energy level easily finds paths leading downward.

 \subsubsection{Energy maps for \texorpdfstring{$N$}{N}=5}

 The full map of the five (and higher) dimensional hypercubes is so crowded that from this point on we display only the lowest energy states and their connections. 
 
\begin{figure}[h]
    \centering
\begin{tikzpicture}[scale=0.45]
\useasboundingbox (-3,-3) rectangle (3,3);
\foreach \i in {30,150,210,330} {
    \filldraw[black] ({3*cos(\i)},{3*sin(\i)}) circle (6pt);
}
\filldraw[Cyan] (0,{3*sin(30)}) circle (6pt);
\filldraw[Cyan] (0,{3*sin(330)}) circle (6pt);
\foreach \i in {90,165,195,270} {
    \filldraw[Cyan] ({1.5*cos(150)},{3*sin(\i)}) circle (6pt);
    \filldraw[Cyan] ({1.5*cos(30)},{3*sin(\i)}) circle (6pt);
}
\draw[black,line width=0.1mm] ({3*cos(150)}, {3*sin(150)}) -- ({3*cos(150)}, {3*sin(210)});
\draw[black,line width=0.1mm] ({3*cos(30)}, {3*sin(30)}) -- ({3*cos(330)}, {3*sin(330)});
\draw[black,line width=0.1mm] ({3*cos(150)}, {3*sin(150)}) -- ({1.5*cos(150)},3) -- (0, {3*sin(30)}) -- ({1.5*cos(30)},3) -- ({3*cos(30)}, {3*sin(30)});
\draw[black,line width=0.1mm] ({3*cos(150)}, {3*sin(150)}) -- ({1.5*cos(150)},{3*sin(165)}) -- (0, {3*sin(30)}) -- ({1.5*cos(30)},{3*sin(165)}) -- ({3*cos(30)}, {3*sin(30)});
\draw[black,line width=0.1mm] ({3*cos(210)}, {3*sin(210)}) -- ({1.5*cos(150)},{3*sin(270)}) -- (0, {3*sin(330)}) -- ({1.5*cos(30)},{3*sin(270)}) -- ({3*cos(330)}, {3*sin(330)});
\draw[black,line width=0.1mm] ({3*cos(210)}, {3*sin(210)}) -- ({1.5*cos(150)},{3*sin(195)}) -- (0, {3*sin(330)}) -- ({1.5*cos(30)},{3*sin(195)}) -- ({3*cos(330)}, {3*sin(330)});
\node[draw] at (-3,-3) {a};
\end{tikzpicture}
\hspace*{15mm}
%
\begin{tikzpicture}[scale=0.45]
\useasboundingbox (-3,-3) rectangle (3,3);
\Poligon{1,0,1,0,1,0}{30}{3}{1}
\filldraw[black] (0,0) circle (6pt);

\foreach \i in {0, ..., 2} {
    \Conn[black]{3}{-30+\i*120}{0}{0}
}
\node[draw] at (-3,-3) {b};
\end{tikzpicture}
\hspace*{15mm}
%
\begin{tikzpicture}[scale=0.45]
\useasboundingbox (-3,-3) rectangle (3,3);
\Poligon{0,0,0,0,0,0,0,0,0,0}{18}{3.5}{1}
\Poligon{1,1,1,1,1,1,1,1,1,1}{18}{2}{0}
\filldraw[Cyan] (0,0) circle (6pt);
\foreach \i in {0,...,9} {
    \Conn[black]{3.5}{\i*36+18}{2}{\i*36+36+18}
    \Conn[black]{2}{\i*36+36+18}{3.5}{\i*36+2*36+18}
}
\foreach \i in {0,...,4} {
    \Conn[black]{0}{0}{2}{\i*36*2-18}
}
\node[draw] at (-3,-3) {c};
\end{tikzpicture}
     \caption{Phase space map of three classes for $N=5$. Only the ground states (black) and some of the first excited states (blue) are shown, with (a) corresponding to Class 3, (b) to Class 6, and (c) to Class 4 in Table~\ref{tab:Table2}. }
     \label{fig:N=5_decagon}
\end{figure}
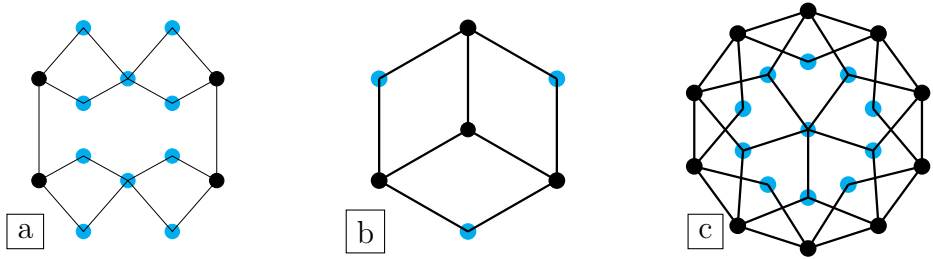

 The first class of configurations for $N=5$ is the usual ferromagnet that does not bring any novelty. Class 2 is similar, with a shift in the spectrum upwards. In Class 3 we find four ground states, with two ground states directly linked and the other two just the reflected image, with the two groups connected by 1st excited states, Fig.~\ref{fig:N=5_decagon}a. Similarly, the eight ground states in Class 6 are organized in two groups of four, the two groups mirror images of each other again, Fig.~\ref{fig:N=5_decagon}b, where only one of the two groups is shown. In these cases the two halves of the phase space are clearly distinguished.

 The most interesting Class for $N=5$ is the fourth. As can be seen in Table~\ref{tab:Table2}, it has a very simple and symmetric energy spectrum with ten ground states. The arrangement of these ground states and the first excited states are exhibited in Fig.~\ref{fig:N=5_decagon}c. The ground states form a decagon whose every vertex is connected to two first excited states. (The ten second excited states form a similar structure, and there are bonds linking the two decagons, but we refrain from showing the full figure, because it would be too crowded and hard to grasp.)
 
\begin{figure}[h]
    \centering
\begin{tikzpicture}[scale=0.45]
\useasboundingbox (-3,-3) rectangle (3,3);
\Poligon{0,0,0,0,0}{18}{3}{1}
\node[draw] at (-3,-3) {a};
\end{tikzpicture}
\hspace*{15mm}
\begin{tikzpicture}[scale=0.45]
\useasboundingbox (-3,-3) rectangle (3,3);
\Poligon{0,0,0,0,0}{18}{3}{0}
\foreach \i in {0,...,4} {
    \Conn[black]{3}{\i*72+18}{3}{\i*72+144+18}
}
\node[draw] at (-3,-3) {b};
\end{tikzpicture}
    \caption{Two symmetric members of the $N=5$ Class 4 explaining the five-fold symmetry of the energy map. Again, only the negative bonds are shown, the positive ones are left understood.}
    
    \label{fig:N=5_pentagon_and_star}
\end{figure}
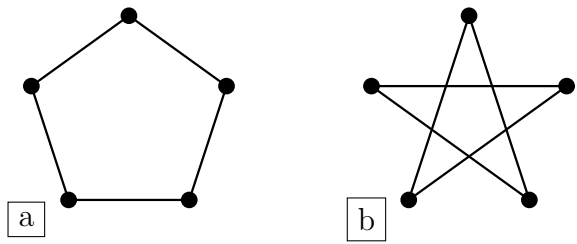

 It may be surprising to see such a highly symmetric map arising from the asymmetric distribution of positive and negative links shown in Fig.~\ref{fig:N=5_configurations}. We have to realize, however, that the representation in that figure is just the most economic one, in that it contains the least number of negative links. Through a few steps of gauge transformations and permutations one can show that an equivalent representation of this class can be given either by a pentagon or a five-pointed star of negative links, the others being positive, as shown in Fig.~\ref{fig:N=5_pentagon_and_star}. This explains the five-fold symmetry of the energy map. It is also a demonstration of how highly symmetric energy landscapes  can be hidden behind the curtain of gauge transformations.  

 This particular Class was regarded as "archetypally" spin glass-like by~\cite{Caliri1985}. They pointed out that it contained precisely the same number of positive and negative bonds, which is certainly true for the pentagon and the five-point star representations, but not for the other gauge transformed variants.
 
 These members of Class 4 display some further interesting properties. If we change the signs of each bond in Fig.~\ref{fig:N=5_pentagon_and_star}a, we get Fig.~\ref{fig:N=5_pentagon_and_star}b, and vice versa. So the energy spectrum of either one of them is the same as that of the negative of the other, while on the other hand they are gauge and permutation equivalent, so have the same spectrum. Therefore this spectrum must be symmetric to the origin, which it is indeed, see Class 4 in Table~\ref{tab:Table2}. The two representatives of Class 4 shown in Fig.~\ref{fig:N=5_pentagon_and_star} are then examples of sign-symmetric signed graphs. It may be interesting to identify configurations with a similar property for higher $N$ values.

 Another remarkable property of this class is that the cube of these interaction matrices $J$ is proportional to the matrix itself: $J^3~=~5J$.

 As an aside we may also note that the graphs in Fig.~\ref{fig:N=5_pentagon_and_star} contain no pure ferromagnetic or pure antiferromagnetic triangle. This means the Ramsey number $R(3,3)$ must be larger than 5. In fact, it is 6 \cite{Greenwood1955}.

 \subsubsection{Energy maps for \texorpdfstring{$N$}{N}=6}

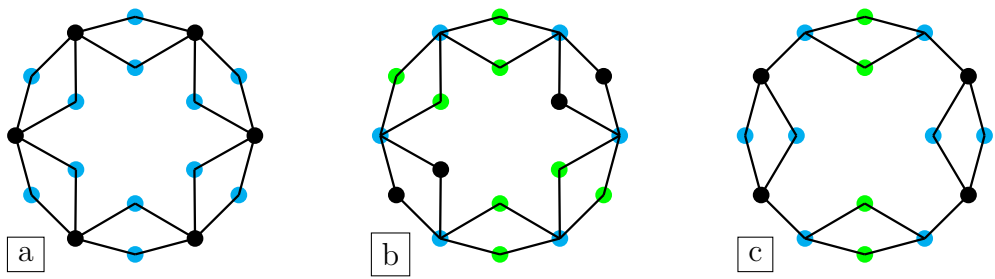
\begin{figure}[h]
\begin{center}
\begin{tikzpicture}[scale=0.45]
\useasboundingbox (-3.5,-3.5) rectangle (7,4);
\Poligon{1,0,1,0,1,0,1,0,1,0,1,0}{30}{3.5}{1}
\Poligon{1,1,1,1,1,1}{30}{2}{0}
\foreach \i in {0,...,5} {
    \Conn[black]{3.5}{\i*60}{2}{\i*60+30}
    \Conn[black]{2}{\i*60+30}{3.5}{\i*60+60}
}
\node[draw] at (-3.2,-3.5) {a};
\end{tikzpicture}
\hspace*{-2mm}
\begin{tikzpicture}[scale=0.45]
\useasboundingbox (-3.5,-3.5) rectangle (7,4);
\Poligon{2,1,2,1,0,1,2,1,2,1,0,1}{90}{3.5}{1}
\Poligon{2,2,0,2,2,0}{90}{2}{0}
\foreach \i in {0,...,5} {
    \Conn[black]{3.5}{\i*60}{2}{\i*60+30}
    \Conn[black]{2}{\i*60+30}{3.5}{\i*60+60}
}
\node[draw] at (-3.2,-3.5) {b};
\end{tikzpicture}
\hspace*{-2mm}
\begin{tikzpicture}[scale=0.45]
\useasboundingbox (-3.5,-3.5) rectangle (3.5,4);
\Poligon{2,1,0,1,0,1,2,1,0,1,0,1}{90}{3.5}{1}
\Poligon{2,1,2,1}{90}{2}{0}
\foreach \i in {0,...,3} {
    \Conn[black]{3.5}{\i*90+60}{2}{\i*90+90}
    \Conn[black]{2}{\i*90+90}{3.5}{\i*90+120}
}

\node[draw] at (-3.2,-3.5) {c};
\end{tikzpicture}
     \caption{Phase space map of Class 6 (a), Class 11 (b) and Class 9 (c) for $N=6$ (black: ground state, blue: 1st excited state, green: 2nd excited state).}
     \label{fig:en_N=6}
\end{center}
\end{figure}

 The first few classes do not show remarkable features: we have doubly degenerate ground states which are separated from each other by a barrier, which is becoming lower and lower as we add more negative bonds. 
 This picture changes in Class 6,~9, and 11 where we can observe how the ground states start to divide the phase space among themselves. The corresponding energy maps are shown in Fig.~\ref{fig:en_N=6}. As can be seen, the ground states have their small basins of attraction separated by low, one- or two energy unit barriers. At this small $N$ one could not expect to find deep energy valleys, but it is not hard to imagine that with increasing $N$ the fragmentation of phase space continues and the barriers grow. In Class 9 and 11 we can also see that the ground states are organized in small families, within which the barriers are lower than between them. We regard these cases as the appearance of the fragmentation of phase space in an embryonic form. The fast growing number of cases make it very hard to analyze the classes in a comparable detail for larger $N$'s, but we are sure that with $N$ growing more and more branches of the ultrametric tree would show up with higher and higher barriers between them.
 
 Classes 7, 8, 9, and 10 have symmetric spectra w.r.t. the origin, which means that their interaction matrices are again sign-symmetric. Also, the trace of their cubes is zero, moreover for Class 10 $J^3$ is again proportional to $J$: $J^3=-5J$.

\begin{figure}
     \centering
\begin{tikzpicture}[scale=0.45]
\useasboundingbox (-4,-4) rectangle (8,8);
\Poligon{0,0,0,0,0,0}{90}{4}{0}
\Poligon{0,1,0,1,0,1}{90}{2}{1}
\Poligon{1,1,1,1,1,1}{0}{3}{0}
\filldraw[Cyan] (0,0) circle (6pt);

\foreach \i in {0, ..., 5} {
    \Conn[black]{4}{90+\i*60}{3}{60+\i*60}
    \Conn[black]{3}{60+\i*60}{4}{30+\i*60}
}
\foreach \i in {0, ..., 2} {
    \Conn[black]{3}{60+\i*120}{2}{90+\i*120}
    \Conn[black]{3}{120+\i*120}{2}{90+\i*120}
    \Conn[black]{4}{30+\i*120}{2}{30+\i*120}
    \Conn[black]{2}{90+\i*120}{0}{0}
}
\Conn[thick,dotted,gray]{2}{330}{2}{90}
\Conn[thick,dotted,gray]{2}{210}{2}{90}
\Conn[thick,dotted,gray]{2}{210}{2}{330}
\Conn[thick,dotted,gray]{2}{90}{4}{30}
\Conn[thick,dotted,gray]{2}{330}{4}{30}
\Conn[thick,dotted,gray]{2}{90}{4}{150}
\Conn[thick,dotted,gray]{2}{210}{4}{150}
\Conn[thick,dotted,gray]{2}{90}{4}{90}
\Conn[thick,dotted,gray]{4}{30}{4}{90}
\Conn[thick,dotted,gray]{4}{150}{4}{90}
\Conn[thick,dotted,gray]{4}{150}{4}{210}
\Conn[thick,dotted,gray]{4}{210}{2}{210}
\Conn[thick,dotted,gray]{4}{210}{4}{270}
\Conn[thick,dotted,gray]{4}{270}{2}{210}
\Conn[thick,dotted,gray]{4}{270}{4}{330}
\Conn[thick,dotted,gray]{4}{330}{2}{330}
\Conn[thick,dotted,gray]{4}{270}{4}{330}
\Conn[thick,dotted,gray]{4}{330}{4}{30}

\Conn[thick,dotted,gray]{4}{270}{2}{330}

\end{tikzpicture}
     \caption{Class 10 for $N=6$. Note that this is not the full map, but only the front view of what is a 3-dimensional body. The gray dotted lines are the edges of the icosahedron, which are just to help to recognize the underlying body, but the ground states they connect are not first neighbors in phase space.}
     \label{fig:N6-Class10l}
 \end{figure}
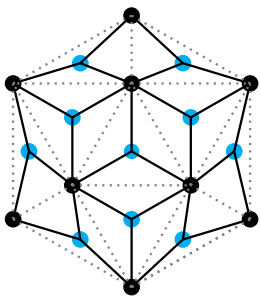

 Class 10 has by far the most interesting energy map: each of its 12 ground states is connected to 5 first excited states, while each of the 20 first excited states is connected to 3 ground states. This can be visualized as the ground states sitting at the vertices of an icosahedron and the excited states at the center of the triangular faces, see  Fig.~\ref{fig:N6-Class10l}. Since the ground states are not directly connected to each other, the edges of the icosahedron are missing. Considering that these spin states are occupying the vertices of a six-dimensional hypercube the emergence of an icosahedron may at first sight seem surprising. 

\subsection{Local minima and isolated clusters for energies higher the the ground state}

The energy landscape of spin glasses displays not only ultrametrically organized ground states, but also "valleys" lying at higher energies. We probed into whether these features show up at the small $N$'s that we are studying here. To this end we looked for local minima and saddle points and also more extended isolated clusters of states that are surrounded by higher energy states. 

\begin{figure}[h]
    \centering
\begin{tikzpicture}[scale=0.45]
\useasboundingbox (0,0) rectangle (7,11);
\foreach \i in {1, ..., 6} {
   \draw[black,line width=0.1mm] ({\i}, 1) -- ({\i}, 10);
   \draw[thick,dotted] ({\i}, 10) -- ({\i}, 11);
   \draw[thick,dotted] ({\i}, 0) -- ({\i}, 1);
   \foreach \j in {1, ..., 10} {
      \draw[black,line width=0.1mm] (1, {\j}) -- (6, {\j});
      \draw[thick,dotted] (6, {\j}) -- (7, {\j});
      \draw[thick,dotted] (0, {\j}) -- (1, {\j});
      \pgfmathsetmacro\mynum{\i+\j}
      \ifodd\mynum\relax
         \filldraw[Cyan] (\i,\j) circle (6pt);
      \else
         \filldraw[green] (\i,\j) circle (6pt);
      \fi
   }
}
\end{tikzpicture} \hspace*{10mm}
\includegraphics[width=0.4\textwidth, trim={0.cm 2.cm 0.cm 0.cm}, clip]{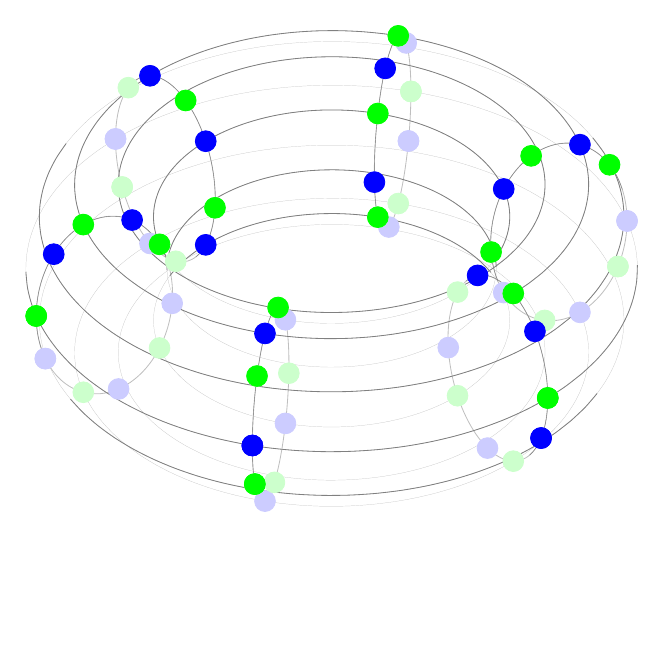}
    \caption{Isolated 60-cluster in $N=8$. The dotted lines on the left figure represent connections to the opposite side, forming the surface of a torus (see right figure). Each spin state is connected to 4 others inside the isolated clusters, and 4 other connection are going 'outside'. The minimal wall to the outside is 1 energy level high for each spin state.}
    \label{fig:isolatedN=8_60_cluster}
\end{figure}
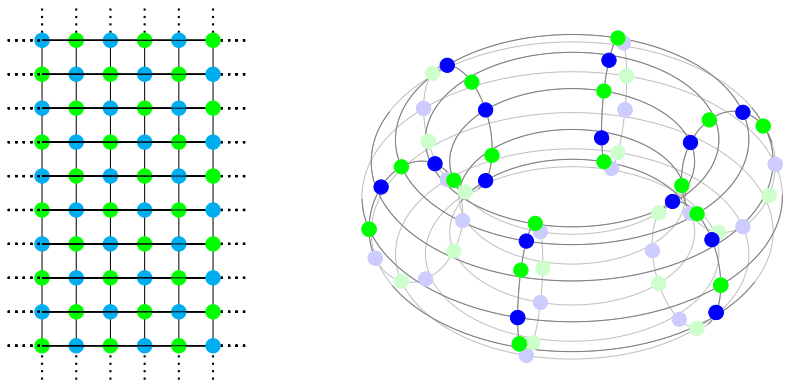

Our findings are as follows. There are no local minima or isolated clusters up to and including $N=5$. For $N=6$ we find three classes with 2, 4 resp. 8 isolated local minima: single spin states whose surrounding states are one step higher in energy. For $N=7$ there is no local minimum or isolated cluster above the ground state(s). 

Things become interesting at $N=8$ where different types of isolated cluster start to develop. In addition to the single-point local minima, more extended ones appear. We identified 55 classes with extended clusters. Of these 6 classes have a single large extended cluster consisting of 60, 42, 28, 24, 24 and 16 spin states, respectively. In 45 classes we found two mirror symmetric clusters the largest ones containing  $2\times18$, $2\times17$, in three classes $2\times12$ and in 4 classes $2\times10$ states. In addition, we found 4 classes having 4 clusters each, with sizes $4\times4$ in one of the classes, and $4\times2$ in three cases. 

At $N=8$ there are 65 classes with multiple (more than 2) ground states. As the total number of invariance classes is 219, we can see that roughly half of them contain multiple ground states and/or extended clusters above the ground states with the states belonging to the cluster having lower energy than their surrounding. It is reasonable to expect that at higher $N$'s these structures will further proliferate.

The full display of all these isolated clusters would take up too much space, so we present here just the most interesting one. This cluster consists of 60 alternating first and second excited states which are sitting on the surface of a torus! The rectangle in Fig.~\ref{fig:isolatedN=8_60_cluster} is a planar representation of this torus where periodic boundary conditions are meant. The spin states neighboring the ones on the surface of the torus have one or two unit higher energy. The torus goes around in phase space, but the states belonging to it do not exhaust the whole set of first and second excited states, so one can walk around it when going from the ground states to the higher energy states without necessarily stepping into this shallow minimum. 

For $N=9$ there are 2 classes with clusters of size 2 (in one of them there are 4 such clusters, in the other 2 clusters), and there are single-state local minima in 75 classes. These numbers are to be compared with the total number of classes 1625 for $N=9$.

Note the marked difference between the even and odd $N$ cases: the even $N$ classes tend to be richer in detail. 

We also observed that these extended objects tend to appear in those classes where the ground state is just two-fold degenerate, but the energy levels one or two steps higher are crowded with a large number of states.

A final remark on the extended objects appearing in these energy maps: The icosahedron popping up in one of the ground state maps at $N=6$ or the torus at $N=8$ are constituted by a set of spin vectors, that is some vertices of a $6-$, resp.   $8-$dimensional cube. The role of the Hamiltonian is merely to designate the spin vectors that belong to a given energy. The icosahedron, the torus and all the other objects exhibited in this paper are simply the intersections of the given energy surface with the hypercube, but these objects are there in the cube, waiting to be picked out.

Concerning the icosahedron hiding in the six-dimensional hypercube, it was a surprise to us, but later search of the literature revealed that its existence had been known to crystallographers since 1984~\cite{Kramer1984}. We have not found reference to the torus in the eight-dimensional cube.

\section{Summary}
We have investigated all the SK spin glass samples up to $N=9$ vertices and classified them into isospectral classes with the help of the gauge- and permutation symmetries. In the course of this work we made some contact with the theory of signed graphs. We have the impression that despite the evident overlap between the problem of signed graphs and that of spin glasses not all the mutual advantages have been fully exploited. For example, it would be interesting to know which are the properties of the interaction matrix $J$ that determine the fragmentation of phase space, or the emergence of such surprising structures like e.g. the torus in one of the $N=8$ classes. There is virtually no discussion of spin glasses without invoking the picture of a rugged energy landscape, but we have never heard such a feature as the torus being mentioned. 

It is obvious that the invariance class determines the spin glass spectrum, but we do not know whether the reverse statement is true. We intend to return to this question in future work. 

It is often pointed out that spin glasses possess equilibrium states, or ground states that are in no obvious symmetry relation with each other. This is certainly true if one draws the interactions from a continuous distribution and averages over them. It is nothing short of a miracle that the ultrametric symmetry shines through even after all this. We used discrete couplings and small sizes with no averaging. In this setup it is natural that the energy maps we find reflect the symmetries of the interaction matrix. What can sometimes obfuscate the picture is how these symmetries may be hidden among the gauge invariant versions of the interaction.   

Having surveyed all the possible cases of the interaction matrix we learned that apart from the purely ferromagnetic case and a few of its close neighbors the fragmentation of phase space is a general feature of these systems; if the ground states are not degenerate then the interesting features move up to higher energies. 

In a way, there is nothing extraordinary about this: we are dealing with a non-convex energy landscape.  What is remarkable is that such a toy model as the SK Hamiltonian is able to reflect a universal feature of Nature. Whether we look at the atomic nuclei, the atoms, molecules, living beings, or ecosystems we see the same pattern: a few kinds of building blocks with competing interactions arrange themselves into a number of stable structures that act as attractors for the unstable and metastable ones. But the essential point is that these attractors are, as a rule, not unique. The picture we have found in our small spin glass samples is the appearance of multiple equilibria~\cite {Parisi_Nobel} in an embryonic form. The main message is how soon these equilibria appear and how soon they start to arrange themselves into families.



.


\section*{Acknowledgements}
We are obliged to Matteo Marsili who five years ago shared with us a code that formed the kernel of the much extended code we have used in this work. We thank Risi Kondor, László Lovász and Tamás Temesvári for helpful comments, but none of them have read the manuscript and are in no way responsible for any mistakes.

\bibliographystyle{unsrt}
\bibliography{sample}

\end{document}